T.A. Sargsian, P.A. Mantashyan[1], and D.B. Hayrapetyan

# Quantum Dot behaviour under the influence of non-resonant structured laser beams

**Abstract:** The study investigates and compares the impact of intense, non-diffractive, non-resonant structured laser beams with various intensity profiles on the properties of InAs/GaAs cylindrical quantum dot. The comparative study demonstrates that the different structured beams, having unique symmetries and hence properties, have significant effects on the confinement potentials and electron probability densities of the quantum dot. It is shown that the system is most sensitive to the intensity and peak position changes of the zeroth order Bessel beam.

Drastic changes in the dressed confinement potentials and electron probability densities under the impact of the abovementioned beams were demonstrated, which can lead to the usage of the following method in such applications, where the precise manipulation of the charge carrier location is required. This study provides new insights into the role of structured laser fields in quantum dot systems, offering possibilities for their use in advanced nanophotonics and quantum information technologies.

## 1 Introduction

Quantum dots (QDs) are semiconductor nanocrystals that confine excitons in all three spatial dimensions, resulting in unique, size-dependent electronic and optical properties. These tuneable characteristics have made QDs ideal candidates for a wide range of optoelectronic applications [1], including light-emitting diodes [2], lasers, and advanced quantum computing devices [3]. Over the past few decades, extensive research has focused on examining the electronic and optical characteristics of QDs, particularly regarding variations in size, shape, and morphological structure. Studies have shown that the properties and behaviour of QDs can be precisely modified, controlled, and enhanced by applying external fields, such as electric [4], magnetic [5], and laser fields [6-13].

One promising area of investigation is the use of laser fields to manipulate excitation states in QDs. Previous studies [7] have demonstrated that tailored laser pulses can selectively control biexciton states, setting the stage for potential quantum information applications [8]. Moreover, recent findings [9] indicate that a single continuous-wave infrared laser can simultaneously trap and excite individual colloidal QDs through two-photon absorption, eliminating the need for separate excitation and trapping sources in nanoscale experiments. Other research works [10,11] has shown that Mollow spectra at high Rabi splittings exhibit linewidth broadening due to excitation-induced dephasing. Nonetheless, resonant excitation remains the most effective route to achieving nearly Fourier transform-limited photon sources, enabling single-photon and cascaded two-photon emissions via Mollow triplet sidebands. While extensive studies have observed excitonic state excitation, manipulation, and optical trapping of QDs using Gaussian laser fields, there remains a significant gap in understanding how structured light beams can further enhance these processes.

Recently researchers have begun to investigate the effects of structured light beams on QDs, which present new opportunities for manipulating quantum states [12-15]. For instance, Holtkemper et al. [12] examined the selection rules governing the excitation of QDs using spatially structured light beams, highlighting their potential for reconstructing higher excited exciton wave functions. Their findings demonstrate that tailored optical fields can selectively manipulate exciton states, thereby enhancing our understanding and control of QD dynamics. In a related study [13], authors explored spatially structured transparency and the transfer of optical vortices through four-wave mixing in a QD nanostructure. Furthermore, in 2023 Sargsian et al. [14] investigated the effects of Gaussian and Bessel laser beams on the linear and nonlinear optical properties of vertically coupled cylindrical QDs. Structured beams, particularly Bessel beams, provide unique advantages over traditional Gaussian beams due to their self-healing and non-diffracting properties, making them highly suitable for deep optical manipulation [16,17], lithography [18,19], and optical induction [20,21].

Notably, Sargsian et al. [14] presented the first observation of QDs under the influence of non-resonant Bessel beams, marking a new approach in the field. The study reported that a vertically coupled QD confinement potential could be adjusted by the intense non-resonant Bessel beam through a laser-dressing parameter or by shifting the laser beam's peak position. The initially symmetrical character of the confining potential leads to the symmetric distribution of the wave functions between the neighboring QDs. Moreover, at high intensities, this adjustment introduced a third local minimum in the interdot barrier region, significantly impacting electron tunnelling. However, the initial symmetry was disrupted with further dislocation of the Bessel beam peak, potentially inducing anticrossing effects between energy levels. Importantly, the Bessel beam modified the system's selection rules, allowing previously forbidden transitions to occur. This research highlights the potential of Bessel beams to en-

[1]**Corresponding author: Paytsar Mantashyan**, A.B. Nalbandyan Institute of Chemical Physics of NAS RA, Yerevan, Armenia; paytsar.mantashyan@rau.am; 0000-0001-6724-4868

**Tigran Sargsian**: A.B. Nalbandyan Institute of Chemical Physics of NAS RA, Yerevan, Armenia; tigran.sargsian@rau.am; 0000-0001-6594-6460

**David Hayrapetyan**: A.B. Nalbandyan Institute of Chemical Physics of NAS RA, Yerevan, Armenia; david.hayrapetyan@rau.am; 0000-0001-9829-4702 able novel methods for optical manipulation and control of quantum nanostructures, opening further possibilities for advancements in quantum and optoelectronic technologies.

Similarly, the researchers have examined the impact of intense laser Bessel beams on excitonic complexes within ellipsoidal QDs [15], demonstrating that the unique properties of Bessel beams can lead to notable changes in exciton dynamics and stability. It is essential to note that in these recent studies [14, 15], the Bessel beams were non-resonant with respect to the quantum transitions of the QD systems, modifying the confinement potential and resulting in additional effects, such as electron tunnelling.

In addition to Bessel beams, various types of structured beams, each with unique propagation and phase characteristics, have been developed and are being widely used in different applications. Among these, Airy beams stand out due to their ability to exhibit curved, self-accelerating trajectories while resisting diffraction over considerable distances, making them valuable in applications such as particle manipulation and optical micromachining [22,23]. The self-healing nature of Airy beams further contributes to their robustness, allowing them to maintain their integrity even after encountering obstacles [24]. Another notable category of structured beams are the Mathieu beams, which possess elliptic or hyperbolic intensity profiles, allowing versatile control over their shape, making them particularly suitable for applications requiring asymmetrical focusing [25]. Vortex beams, characterized by their helical phase fronts and ability to carry orbital angular momentum, represent another critical class of structured beams, enabling precise manipulation of particles at the nanoscale and proving indispensable in fields such as quantum optics and optical communication [26,27]. The structured nature of these beams allows researchers to exploit their unique characteristics for enhanced control over light-matter interactions, paving the way for advancements in quantum optics and photonic applications [28].

Building on the findings obtained in previous research on the behavior of QDs under the influence of Bessel beams [14,15], this study aims to extend the investigation to explore the effects of non-resonant structured beams, such as non-diffractive high-order Bessel, Airy, and Mathieu laser beams, on InAs/GaAs cylindrical QDs. The study demonstrates that the different structured beams having unique symmetries and hence properties have significant effects on the confinement potentials and electron probability densities of the QD.

The paper is organized as follows: Section 2 introduces the theoretical model for the problem and the proposed experimental setup. Section 3 presents a detailed analysis of the results, divided into five subsections that address the electronic properties of the system both in the absence of laser fields and under the influence of Gaussian, Bessel, Airy, and Mathieu intense laser beams. A final subsection compares the energetic properties across these conditions. Lastly, Section 4 summarizes the primary conclusions of the study.

## 2 Method

### 2.1 Floquet Formalism

As it was mentioned above, the system under consideration is a cylindrical InAs/GaAs QD. Both in axial and radial directions confinement is modeled with Woods-Saxon potential model:

$$V(\rho,\varphi) = U_0 - \frac{U_0}{1+\exp\left(\frac{\rho-\beta_{rad}}{\gamma_{rad}}\right)},$$

$$V(z) = U_0 - \frac{U_0}{1+\exp\left(\frac{|z|-\beta_{ax}}{\gamma_{ax}}\right)}, \quad (1)$$

$$V(\vec{r}) = V(z) + V(\rho,\varphi),$$

where $U_0$ is the height of potential barrier, $\beta_{ax}$ and $\beta_{rad}$ are the half-widths, $\gamma_{ax}$ and $\gamma_{rad}$ are confinement barrier slopes in the axial and radial directions, respectively. The Wood-Saxon potential is validated by its ability to accurately represent the smoothing of an initially rectangular confinement potential caused by self-induced polarization. This transformation, resulting from the image charge method [29], modifies the sharp potential profile into a smoother form. The resulting potential profile is well-approximated by the Wood-Saxon potential, achieving high fidelity in fitting.

As a starting point, let us present the three-dimensional Schrödinger equation. For the system under the impact of laser beam one can write [30-32]:

$$i\hbar\frac{\partial \Psi(\vec{r},t)}{\partial t} = -\frac{\hbar^2}{2m^*}\nabla^2\Psi(\vec{r},t) + V(\vec{r}+\vec{\alpha}(\vec{r},t))\Psi(\vec{r},t), \quad (2)$$

where $\vec{r}$ and $m^*$ are the electron radius-vector and effective mass, respectively, $V(\vec{r}+\vec{\alpha}(\vec{r},t))$ is the laser dressed confinement potential function, with:

$$\vec{\alpha}(\vec{r},t) = \alpha_0 \cdot \alpha(\vec{r})\sin(\Omega t)\cdot \hat{u}, \quad (3)$$

where $\alpha_0 = \frac{eF_0}{m^*\Omega^2}$ is laser-dressing parameter, $e$ is the electron charge magnitude, $F_0$ is the field strength, $\Omega$ is the laser frequency, $\hat{u}$ is the polarization unit vector. Opposed to the common approach [33,34] we are not applying here the dipole approximation as the laser fields considered in this work vary significantly in the physically important region. Hence, we introduce the multiplier $\alpha(\vec{r})$, which describes the incident laser spatial profile [35]. The form of $\alpha(\vec{r})$ for each particular type of laser beam profile will be brought in the corresponding subsection of Section 3.

The potential $V(\vec{r}+\vec{\alpha}(\vec{r},t))$ is a periodic function of time oscillating at frequency $\Omega$. In the case of high-frequency limit $\Omega\tau \gg 1$, where $\tau$ is the characteristic transit time of the electron in the structure, the electron motion is dominated by the oscillation under the laser field and consequently the electron "sees" the averaged laser-dressed potential, given by the zero-order term of the Fourier expansion of $V(\vec{r}+\vec{\alpha}(\vec{r},t))$ [33]:

$$\langle V(\vec{r},\alpha_0)\rangle = \frac{1}{T}\int_0^T V(\vec{r}+\vec{\alpha}(\vec{r},t))dt. \quad (4)$$

By applying the Floquet approach [32], for the zeroth Floquet component, the system reduces to the time-independent Schrödinger equation:

$$\left[-\frac{\hbar^2}{2m^*}\nabla^2 + \langle V(\vec{r},\alpha_0)\rangle\right]\varphi(\vec{r},t) = E\varphi(\vec{r},t), \quad (5)$$

where $E$ is the particle energy.

In the current work, we investigate the impact of intense, linearly-polarized laser beam traveling in axial direction of the structure. This means that the polarization vector lies in radial plane and one can fully separate axial and radial subsystems. Only ground state will be considered in axial direction with energy $E_{z0}$ and wavefunction $\varphi_{z0}$, while for radial direction the following equation is obtained:

$$\left[-\frac{\hbar^2}{2m^*}\nabla_\rho^2 + \langle V(\rho,\alpha_0)\rangle\right]\varphi_\rho(\rho,t) = E_\rho\varphi_\rho(\rho,t). \quad (6)$$

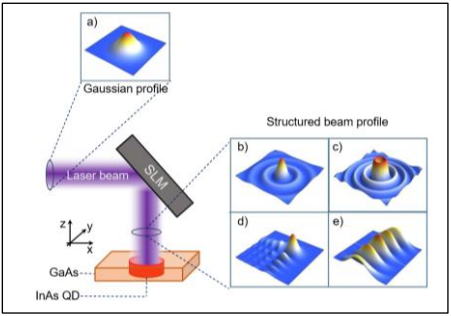

**Fig. 1:** Scheme of cylindrical InAs/GaAs QD irradiation with intense laser field. After entering SLM, initial Gaussian beam (a) changes its profile to either zeroth (b) or first (c) order Bessel, Airy (d) and Mathieu (e) forms.

To transform an initial Gaussian laser beam profile (Figure 1: a inset) into a structured beam profile, various optical elements and techniques can be employed, including spatial light modulators (SLMs) [36], metasurfaces [37,38], and diffractive optical elements (DOEs) [39]. The proposed illumination scheme for InAs/GaAs QDs utilizing structured beams with zeroth (b), first (c) order Bessel, Airy (d), or Mathieu (e) profiles is illustrated in Figure 1.

### 2.2 Material Parameters for InAs/GaAs QD

Here we bring the material parameters for the system under consideration. The QD discussed is made of InAs grown inside GaAs matrix. The material parameters which have been used in the calculations are the following: $m^*=0.023m_0$ and $m^*_{GaAs}=0.063m_0$ are the electron effective masses for InAs and GaAs, respectively, where $m_0$ is the free electron mass, $\varepsilon_r = 15.15$ is the InAs dielectric constant, $E_{g,InAs} = 0.405$ eV and $E_{g,GaAs} = 1.507$ eV are the energy band gaps for InAs and GaAs, respectively [40-43]. We have taken the potential parameters as follows: $\beta_{rad} = 52.5$ nm, $\beta_{ax} = 105$ nm, $\gamma_{rad} = \gamma_{ax} = 3.5$ nm. The value of the depth of the QD potential $U_0$ is, obviously, dependent on the band gap values. The ratio between conduction and valence band discontinuities is based on Ref. [43] and equal to $\Delta E_c / \Delta E_v \approx 1.4047$, with $U_0 = \Delta E_c$.

## 3 Results and discussion

In this section, we analyze the results obtained for various shaped laser beams interacting with the InAs/GaAs cylindrical QD.

### 3.1 QD under the influence of Gaussian beam

The behavior of an InAs/GaAs QD under the influence of an intense Gaussian laser beam is studied. Initially, the QD was analyzed in the absence of an intense laser field to establish a baseline for understanding its natural behavior (Figure 2a-d). In the absence of incident laser field, radial confinement potential has the profile portrayed in Figure 2a. The corresponding density plot for the electron is brought in Figure 2b. As the profiles of both radial and axial confinement potentials are symmetrical in respect to the center of the system, the corresponding probability densities of ground state are also symmetrically distributed around the center of the system (see Figure 2c for the square of the absolute value of radial component and Figure 2d for the same of total wave function in the cylindrical QD).

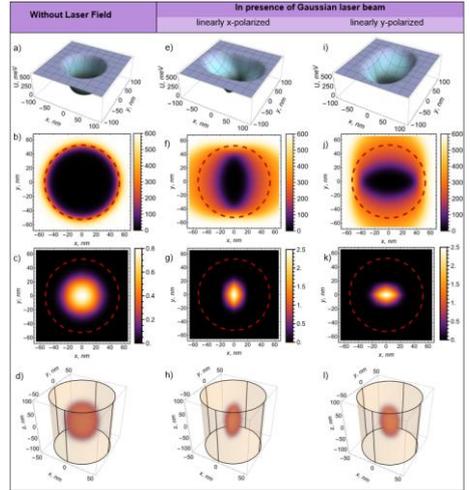

**Fig. 2:** Quantum Dot behaviour in the absence and presence of intense Gaussian beam polarized in (e-h) x or (i-l) y direction. 3D (a, e, i) and density plot (b, f, j) visualizations of radial confinement potential, electron probability density in radial direction (c, g, k) and in 3D (d, h, l).

The investigation of QD was then extended to include the effects of an intense Gaussian beam for both linear x- and y-polarized cases (Figure 2: second and third columns) to understand also the polarization-dependent dynamics of the QD.

When discussing the impact of intense laser field on the cylindrical QD, we will consider 157 nm $F_2$ laser, polarized in the radial direction of the QD. The $\alpha_0$ parameter is taken equal to 35 nm everywhere in the work, if not underlined specifically. First, let us consider Gaussian laser beam, as the initial beam in the absence of axicons and other optical elements will remain in the Gaussian form. For $\alpha(\vec{r})$ in the case of Gaussian beam, one can write:

$$\alpha(x,y) \sim exp\left[-\frac{x^2+y^2}{w^2}\right], \quad (7)$$

where $w$ is the beam waist radius. For our theoretical calculations, the divergence of the beam is chosen as 5 degrees. Figure 2 (2nd and 3rd columns) depicts the effect of Gaussian laser beam on the confinement potential profile and probability densities of the ground state. Second column (Figure 2e-h) corresponds to the x-polarization, while the third one to the y-polarization (Figure 2i-l). It is obvious that in such symmetrical system, the switch between x- and y- polarizations does not lead to any qualitative differences, so the further consideration of polarization switching will be skipped. Shift of the peak position of Gaussian laser beam has a negligible impact on the confinement potential and, hence, on the probability densities in the discussed problem, so it will not be considered in this work [14]. Let us note that this shift can be presented as additional terms $x_{shift}$ and $y_{shift}$ to the $x$ and $y$ coordinates in the corresponding formulas for $\alpha$, i.e. $x \to x + x_{shift}$ and $y \to y + y_{shift}$.

## 3.2 QD under the influence of Bessel beam

As the next step, we will consider the impact of laser beam on cylindrical QD with intensity distribution described by Bessel function. Those beams can be described with the following relation:

$$\alpha(x,y) \sim J_n\left(\frac{2\pi}{\lambda}\sqrt{x^2+y^2}\right), \quad (8)$$

where $J_n$ is the n-th order Bessel function and $\lambda$ is the laser wavelength. Confinement potential profile experiences drastic change under the influence of zeroth order Bessel laser beam. As it is known, Bessel beams offer the advantage of non-diffracting propagation over long distances and can maintain their shape and intensity, unlike Gaussian beams, which spread and lose energy as they propagate.

In this section, we examine the influence of the zeroth- and first-order Bessel beams (shown in the upper left and right insets of Figure 3) on the QD. Starting from specific values of the $\alpha_0$ parameter ($\sim 21$ nm as calculated in Ref. [44]) for the 0th order Bessel beam, a peak forms in the confinement potential profile (Figures 3a and 3b) when the central lobe of the Bessel beam irradiates the center of the QD.

This leads to the separation of probability densities by this peak, which can be seen from Figure 3c and 3d. Shift of the Bessel beam's central lobe by $\beta_{rad}/2$ in x direction from the centre of the system (Figure 3e-h) leads to the shift of confinement potential distortion (Figure 3e and 3f) and probability density distribution (Figure 3g and 3h).

With the means of axicons and other SLM devices, one can obtain higher order Bessel laser beams [45]. Here we will also analyze the impact of first-order Bessel beam on the system under consideration. In contrary to the previous case, the first-order Bessel beam features a central dark spot with null intensity surrounded by alternating bright and dark rings (Figure 3 upper right inset). This circumstance leads to the corresponding appearance of the confinement potential portrayed in Figure 3i and 3h. Eventually, this also forces the electron to be fully localized in the center of the QD, where the dark spot of the beam is incident (see Figure 3k and 3l). When shifting the laser beam peak from the center of the QD, the corresponding confinement pit also moves from the center (as seen from Figure 3m and 3n) "carrying" the electron with itself (Figure 3o and 3p). Hence, the first order Bessel beam can be used as a tool for precise control of charge carrier's localization.

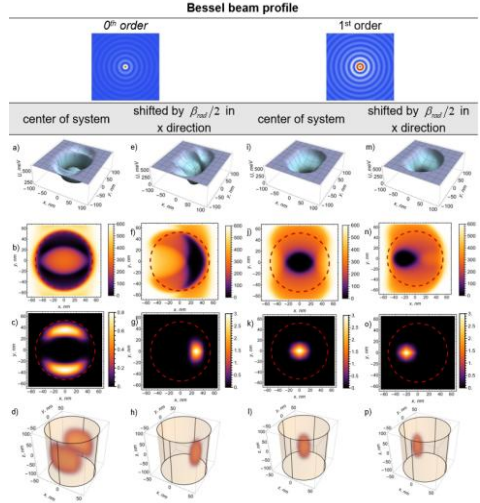

**Fig. 3: Impact of 0th and 1st Order Bessel Beams on a Cylindrical Quantum Dot.** First row: Profiles of the 0th (left) and 1st (right) order Bessel beams. Second row: 3D visualizations of the radial confinement potential under irradiation by 0th (a, e) and 1st order (i, m) Bessel beams. Third row: Density plots of the radial confinement potential for irradiation by 0th (b, f) and 1st order (j, n) Bessel beams. Fourth row: Electron probability density along the radial direction for irradiation by 0th (c, g) and 1st order (k, o) Bessel beams. Fifth row: 3D electron probability density distribution in the radial direction for irradiation by 0th (d, h) and 1st order (l, p) Bessel beams. The Bessel beams were y-polarized, with the peak positioned at the centre of the system (first and third columns) and shifted by a distance of $\beta_{rad}/2$ along the x direction (second and fourth columns).

In addition, to investigate the influence of laser parameter $\alpha_0$ on the modification of QD properties, we analyzed how the depth and shape of the confinement potential profile varied with different values of the laser parameter. Thus, we have demonstrated that by tuning the laser parameter, one can control both the depth and the shape of the confinement potential profile, resulting in a modification of the electron probability density distribution (see Figure 4). Before obtaining the form presented in Figure 3c, the electron probability density distribution is being modified from the initial form of Figure 2c according to the steps described by Figure 4 for $\alpha_0$ parameter values of 14nm (a), 21nm (b), 28nm (c), 31.5nm (d), respectively. As it was discussed under

Figure 2c, one obtains symmetric behavior of electron probability density when any external perturbations are absent.

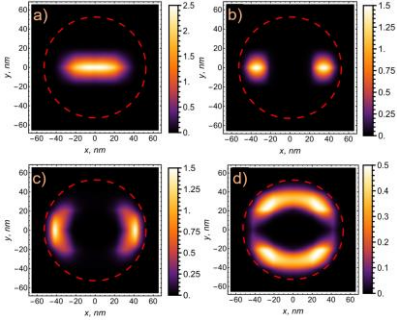

**Fig. 4: The influence of laser parameter α₀ on the modification of QD properties.** Electron probability densities in radial direction for y-polarized 0th order Bessel beam with peak position in center of system for a) 14nm, b) 21nm, c) 28nm and d) 31.5nm values of $\alpha_0$ parameter.

When the system under consideration is being affected with a linearly y-polarized non-diffractive intense laser field with zeroth order Bessel profile, the electron experiences "pressure" from upper and bottom sides of y direction due to the deformation of confinement potential. This leads to the "prolonged" appearance of probability density depicted in Figure 4a. Further increase of the incident beam intensity "slams" the barriers from both sides, which results in division of probability density into two pieces, portrayed in Figure 4b. The more increase in the intensity of beam, the more electron is "smeared" into the right and left walls of QD (Figure 4c) until the deformation of confinement potential eventually "forces" the electron to escape to upper and bottom walls (Figure 4d and 3c). This electron localization "rotation" in XY plane with the increase in intensity has similarities with the effect of the polarization "switching" from x- to y-axis when considering, for example, the situation in Figure 2j and 2g. All of the abovementioned shows that with the means of intense laser radiation one can manipulate the probability density of electron in a manner such as that will be equivalent to having multiple QDs with different geometrical form.

### 3.3  QD under the influence of Airy beam

Another class of beams, which will be considered here, are the so-called Airy beams. Named after the Airy function that describes their shape, those are a type of non-diffracting beams with unique properties such as self-healing and non-dispersive behavior (see left inset in Figure 5). One of the seminal works on Airy beams is Ref. [46], where the authors experimentally demonstrated the generation and propagation of Airy beams using an optical setup involving phase modulation. The unique properties of Airy beams make them attractive for various applications in optics, including optical trapping, micromanipulation, and beam shaping.

For Airy beams, $\alpha(\vec{r})$ can be given in the following form:

$$\alpha(x,y) \sim Ai\left(\frac{x}{\chi_0 w}\right)\exp\left(\frac{a_x}{\chi_0 w}x\right) Ai\left(\frac{y}{\chi_0 w}\right)\exp\left(\frac{a_y}{\chi_0 w}y\right), \quad (9)$$

where $Ai$ is the Airy function of the first kind, $a_x$ and $a_y$ are the decay factors that are positive quantities to ensure the containment of the infinite Airy tail in the x and y directions, and $\chi_0$ is the distribution factor that makes the beam tend to an Airy beam when it is a low value, or a Gaussian beam when it is a high value [47].

In our calculations we choose $\alpha_0 = 8.75$ nm, $\chi_0 = 0.01$ and $a_x = a_y = 0.1$. The unique, multi-peak profile of the Airy beam leaves its footprint on the confinement potential profile (Figure 5a and 5b). In the places of beam peaks, one obtains confinement barriers, which force the charge carrier to be localized in the further position of the QD (Figure 5c and 5d).

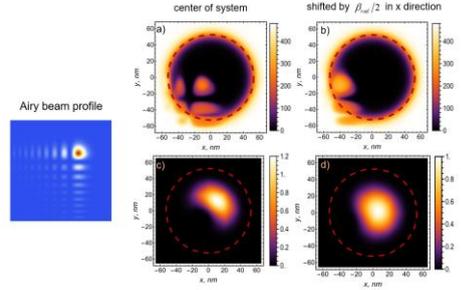

**Fig. 5. Impact of Airy Beam on a Cylindrical Quantum Dot.** (a, b) Density plot visualizations of radial confinement potentials and (c, d) electron probability densities in radial direction for y-polarized Airy beam (left inset) with peak position in center of system (a, c) and shifted by $\beta_{rad}/2$ in x direction (b, d).

### 3.4  QD under the influence of Mathieu beams

Finally, the last case to be investigated in this work are the so-called Mathieu beams. Mathieu beams are a class of non-diffractive laser beams that have a transverse intensity profile described by Mathieu functions (see upper row of Figure 6). These beams are solutions to the paraxial wave equation with elliptical symmetry. Similar to other types of non-diffractive beams, Mathieu beams exhibit some key properties, such as self-healing and good tunability, which make them valuable for applications in various fields, including laser physics, optical trapping and manipulation, microscopy, and beam shaping.

For $\alpha(\vec{r})$ in the case of even and odd Mathieu beams, one can write, respectively:

$$\alpha(\xi,\eta) \sim \begin{cases} ce_r(\eta,q) Je_r(\xi,q) & \text{even}, \\ se_r(\eta,q) Jo_r(\xi,q) & \text{odd}, \end{cases} \quad (10)$$

where $\xi$, $\eta$, $z$ are the elliptical coordinates, which are related to Cartesian coordinates as:

$$x = f \cdot \cosh(\xi)\cos(\eta), \quad y = f \cdot \sinh(\xi)\sin(\eta), \quad z = z,$$

$q = \dfrac{f^2 k_t^2}{4}$ is the ellipticity factor, $f$ is the half focal length in the elliptical-cylindrical coordinate system, $k_t$ is the radial wave number, $ce_r$ and $Je_r$ are even angular and radial Mathieu functions of order r, $se_r$ and $Jo_r$ are odd angular and radial Mathieu functions of order r [48].

Cases with $q=0$ correspond to Bessel beams of the same r-th order [49], so they will not be considered again. We have pictured the radial confinement potentials and electron probability densities in each row of Figure 6 for 0th to 3rd orders of Mathieu beams with $q=15$. Without repeating ourselves, let us note, that all abovementioned regularities in the behavior of confinement potential profiles and, hence, probability densities are applicable to the case of Mathieu beams.

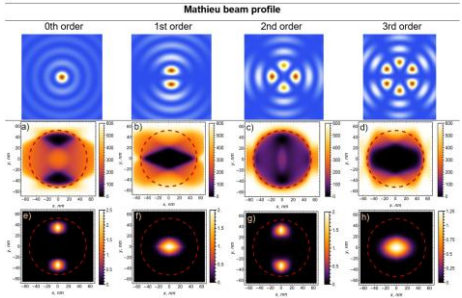

**Fig. 6: Impact of Mathieu Beam on a Cylindrical Quantum Dot.** (upper row) Mathieu beam profiles for 0th, 1st, 2nd and 3rd orders, respectively, and q=15. (a-d) Density plot visualization of radial confinement potential and (e-h) electron probability density in radial direction for x-polarized Mathieu beams (each row corresponds to 0th, 1st, 2nd and 3rd orders, respectively).

### 3.5 QD energetic behaviour under various structured laser beam types

Before this point, we focused on examining the behavior of the electron wave functions in a cylindrical QD under the influence of various shaped laser beams. Now, we proceed to analyze the ground-state energies of the electron. A comparative analysis of these energies is presented in Figure 7, highlighting the effects of different laser beam profiles. The 3D histogram illustrates the dependance of ground state energy on the $\alpha_0$ laser parameter and beam peak $y_{shift}$ shifts for Gaussian (G) and all of the abovementioned types of non-diffractive laser beams: zeroth- and first-order Bessel beams, both unshifted and shifted (B0, B1, B0sh, B1sh); Airy beams (A); and Mathieu beams of zeroth, first, second, and third order (M0, M1, M2, M3). For the reference, ground state energies are brought in the absence of any laser fields (WL).

Increase in the values of $\alpha_0$ laser parameter leads to the increase of the ground state energies (see Figure 7a) as a result of additional confinement potential arising from laser-dressing. However, this increase has different rates for different types of lasers. For instance, with the increase of $\alpha_0$ from 0 (corresponding to the WL case) to 17.5nm, the ground state energy increases by 46.52% for Gaussian beams; 1404.04% and 141.09% for zeroth-order unshifted and shifted Bessel beams, respectively; 89.98% and 113.98% for first-order unshifted and shifted Bessel beams, respectively; 174.24% for Airy beams; and 221.52%, 60.36%, 112.03%, and 21.75% for zeroth-, first-, second-, and third-order Mathieu beams, respectively.

The abovementioned yet again underlines that the system is the most sensitive to the increase of $\alpha_0$ for the zeroth-order Bessel beam due to its pronounced central peak. Shift in the position of the incident laser beam also affects the energetic properties of the system due to the interplay of laser-dressing and size-quantization. Subsequent shifting of the laser peak position from $y_{shift}=26.25$ nm to the center of the QD $y_{shift}=0$ results in increases of 0.04%, 523.86%, -11.22%, 100.33%, 34.99%, 4.85%, 44.26%, and 10.06% for Gaussian, zeroth- and first-order Bessel, Airy, and zeroth-, first-, second-, and third-order Mathieu beams, respectively (see Figure 7b).

As it was stated, the shift in the peak position of Gaussian beam has practically no effect on energetic values due to the diffractive characteristics of the beam. The decrease in the case of first-order Bessel beam can be explained by the nature of the latter: in contrast to other cases the central spot here is dark. Thus, the most distorted confinement part of this type of beam is located near the bright spots, altering the confinement potential in a manner that slops the barriers and deepens the central pit (see Figure 3).

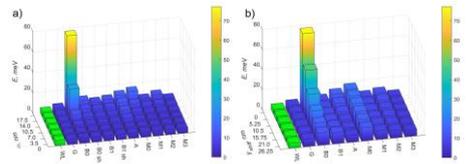

**Fig. 7: 3D histograms of ground state energies in QD under the influence of Gaussian and various non-diffractive structured laser beams depending on $\alpha_0$ parameter.** Ground state energy values are shown for the absence (WL) and presence of Gaussian (G), zeroth- and first-order unshifted and shifted Bessel beams (B0, B0sh, B1, B1sh), Airy beams (A), and zeroth-, first-, second-, and third-order Mathieu beams (M0, M1, M2, M3) at different values of $\alpha_0$ when a) $x_{shift}=y_{shift}=0$, and b) $y_{shift}$ when $\alpha_0=17.5$ nm and $x_{shift}=0$. WL differs in color as it is brought for comparison purpose.

## 4 Conclusion

In conclusion, this comparative study was devoted to the investigation of the impact of intense, non-diffractive, non-resonant structured laser beams with various intensity profiles on the properties of InAs/GaAs cylindrical quantum dot. Different types of laser beams, including Gaussian, Bessel, Airy, and Mathieu beams were examined, and their impact on the electronic properties and confinement potentials of the quantum dot were analyzed.

The study demonstrates that the different laser beams have significant effects on the confinement potentials and electron probability densities of the quantum dot. The zeroth-order Bessel beam, in particular, shows pronounced changes in the confinement potential profile and electron localization. We have also observed that the intensity and peak position of the laser beams play a significant role in modifying the energetic properties of the system.

These results have important implications for the precise control and manipulation of charge carrier localization in quantum nanostructures. The use of intense laser beams, such as Bessel beams, can offer new possibilities for applications that require precise manipulation of charge carrier location. This study provides valuable insights into the potential applications of non-diffractive laser beams in the field of quantum nanostructures.

## 5 Authors' statements

**Research funding:** The research was supported by the Higher Education and Science Committee of MESCS RA (Research project № 24FP-2A029).

**Author contribution:** All authors have accepted responsibility for the entire content of this manuscript and consented to its submission to the journal. TS developed the model code and performed the simulations and interpreted the reported study. PM contributed to the conception of the study, designed and prepared the manuscript with contributions from all co-authors. DH defined the problem of the study and executed the whole procedure of the research.

**Conflict of interest**: Authors state no conflict of interest.

**Data availability statement:** The datasets generated and/or analysed during the current study are available from the corresponding author upon reasonable request.